\documentclass[aps,prb,preprint,groupedaddress,showpacs]{revtex4}

\usepackage{amsmath}
\usepackage{amssymb}
\usepackage{epsfig}
\newcommand {\dr}{{\mathrm d}\mathbf{r}}

\newcommand {\rr}{\mathbf{r}}
\newcommand {\etal}{\begin{itshape}et al\end{itshape}.}

\begin{document}


\title{Solvent mediated interactions close to fluid-fluid phase
separation: microscopic treatment of bridging in a soft core fluid}



\author{A.J.~Archer}
\email[]{Andrew.Archer@bristol.ac.uk}
\author{R.~Evans}
\affiliation{H.~H.~Wills Physics Laboratory, University of Bristol,
Bristol BS8 1TL, UK}
\author{R.~Roth}
\author{M.~Oettel}
\affiliation{Max-Planck-Institut f\"{u}r
Metallforschung, Heisenbergstr.\ 3, 70569 Stuttgart, Germany}
\affiliation{ITAP,
Universit\"{a}t Stuttgart,  Pfaffenwaldring 57, 70569 Stuttgart, Germany}



\date{\today}

\begin{abstract}
Using density functional theory we calculate the density profiles of a binary
solvent adsorbed around a pair of big solute particles. All species interact via
repulsive Gaussian potentials. The solvent exhibits fluid-fluid phase separation
and for thermodynamic states near to coexistence the big particles can be
surrounded by a thick adsorbed `wetting' film of the coexisting solvent phase.
On reducing the separation between the two big particles we find there can be a
`bridging' transition as the wetting films join to form a fluid bridge.
The potential between the two
big particles becomes long ranged and strongly attractive in the bridged
configuration. Within our mean-field treatment the bridging transition results
in a discontinuity in the solvent mediated force.
We demonstrate that accounting for the phenomenon of bridging requires the
presence of a non-zero bridge function in the correlations
between the solute particles when our model fluid is described within a full
mixture theory  based upon the Ornstein-Zernike equations.

\end{abstract}


\maketitle

\section{Introduction}

Big solute particles (e.g.\ colloids) immersed in a solvent of smaller particles
interact with each other by an effective potential which is the
sum of their direct interaction and a solvent mediated (SM) potential. 
Even when the direct interaction consists solely of two--body terms, the SM
potential
contains higher body contributions, of all orders which are determined formally by
integrating out the solvent degrees of freedom. This conceptual framework
yields, in principle, a much simpler effective Hamiltonian which
involves only the coordinates of the big particles. \cite{Likos}
In certain systems 
the two--body term in the SM potential may dominate completely the corresponding
direct interaction. A well--known example is a suspension of big hard--sphere colloids
in a solvent of small hard--spheres. There the SM potential between the colloids is termed
the depletion interaction and this is the only contribution to the effective
potential for separations greater than the big hard--sphere diameter.
\cite{Roland} In the case of a (non--hard) solvent which is
at a state point near to fluid-fluid phase separation, big solute particles
can be surrounded by a thick adsorbed `wetting' film of the coexisting
solvent phase.
\cite{Dietrich} If two such big particles become sufficiently close, there
can be a `bridging transition' as the wetting films surrounding the two big
particles join to form a fluid bridge of the wetting phase -- see for example
Ref.~\onlinecite{BauerPRE2000} and references therein. In wet granular media
these bridging (or capillary) forces lead to strong and very short--ranged
interactions. For small solutes these effective attractions become rather
long--ranged with respect to
the dimension of the solute as is known from tip--substrate interactions in atomic
force microscopy. \cite{Isra} Long--ranged attractive interactions are also
surmised for hydrophobic molecules in water at ambient conditions. \cite{LCW} 
Bridging is also a possible mechanism for driving colloidal flocculation.
\cite{Beysens:JSP99}  

In previous work, \cite{Archer3,Archer5} the wetting of a binary solvent around
a single big particle and the influence of these thick adsorbed films on the
effective SM potential between two big particles, was investigated for a
particular model fluid, namely the generalisation to mixtures of
the Gaussian core model (GCM). \cite{Likos, Archer3,
Archer5, Stillinger, paper3, LangJPCM, paper1, Ard,
Bolhuis, Archer1, Archer2, Finken} A Gaussian potential provides a good
approximation for the effective potential between the centres of mass of
polymers in solution. \cite{Likos, Flory, Dautenhahn}
The approach to calculating the SM potentials was based upon the
theory developed by Roth \etal\ \cite{Roland} -- henceforth referred
to as the `insertion method'. The insertion method works within the framework of
density functional theory (DFT) \cite{Bob}
and uses as input the density profiles
calculated around a {\em single} big particle in order
to calculate the SM potential between a {\em pair} of big particles.
\cite{Archer3,Archer5,Roland} Although the insertion method is formally exact,
in practice one must employ an approximation for the free energy functional of
the mixture of big and small particles. \cite{Roland} For state points near to
coexistence we found thick adsorbed films around the big particles
resulting in long ranged, strongly attractive SM potentials whose range was
determined by the thickness of the wetting film. However,
using the insertion method, we were unable to detect
any direct sign of bridging in the SM potential. \cite{Archer3,Archer5}

The present work can be viewed as going a significant step further than 
Refs.~\onlinecite{Archer3,Archer5}. Here we investigate the same system: two
large solute 
Gaussian particles, immersed in a binary GCM solvent near to phase separation.
However, whereas the previous work used the elegant insertion method, the
present work can be viewed as the `brute-force' approach to the problem.
Using an accurate DFT for the binary GCM solvent of small particles
\cite{Likos, Archer1, Archer2, Archer3, Archer5} we calculate explicitly
the solvent density profiles around a fixed {\em pair} of the big GCM particles,
treating the latter as external potentials, and determine the resulting grand
potential. By repeating this calculation for a range of values of
the separation between the centres of the two big particles we obtain
the SM potential. We find, within the present (mean-field)
DFT approach, that when thick adsorbed films are present
there can be a bridging transition as the separation between the two big
particles is decreased, i.e.\ the formation of a bridging configuration gives
rise to a discontinuity in the derivative of the SM potential.
We believe that this is the first time a non-local DFT has been used to
calculate bridging density profiles and the resulting effective potential
between two particles. 
Bridging has been investigated previously within coarse-grained, local DFT in the
recent study of Stark \etal\ \cite{Stark} for  big hard spherical colloids immersed 
in an isotropic liquid crystal host close to the isotropic-nematic phase boundary.
Similarly, Andrienko \etal\ \cite{AndrienkoetalJCP2004}
calculated bridging density profiles of a solvent adsorbed between a big
colloid and a planar wall using a local DFT.

We also investigate the SM potential between two big GCM particles in a region
of the solvent phase diagram near the binodal but lying
outside the single particle
thin-thick adsorbed film transition line, \cite{Archer5} where a single big
particle does not have a thick adsorbed `wetting' film of the coexisting 
solvent phase around it. Adsorption still influences strongly the SM potential.
We find an analogue of capillary condensation; as the two big particles
become sufficiently close, the composite object is large enough to induce
condensation of the coexisting solvent phase around the {\em pair}
of big particles. This effect is somewhat different from that which can occur
between two big hard-core
particles in a solvent near to coexistence. When a pair of such particles are
sufficiently close, a bridge of the coexisting phase can condense in the gap
between the two big particles, without there being thick `wetting' films
adsorbed on each of the big particles. \cite{BauerPRE2000} In the present
soft core system the strong adsorption
is not confined to the space between the big
particles, rather it extends through the whole region in which the two big
particles are situated. This local condensation also
results in a jump in the SM force
between the two big particles with the SM potential becoming strongly
attractive for small separations.

In the final part of the present work we relate our results for the SM
potential to an approach for calculating the SM potential based upon the mixture
Ornstein-Zernike (OZ) equations. \cite{HM,AmokraneMalherbeJPCM2001,
HendersonPlischkeJCP92} By solving the OZ equations together with a closure
relation one can calculate the various fluid correlation functions. It is well
known that
if one makes a diagrammatic expansion for the fluid correlation functions the
hyper-netted chain (HNC) closure approximation neglects a certain class of
(bridge) diagrams which, taken together, is termed the bridge function.
\cite{HM} 
We show that in order to account for
the phenomenon of bridging of solvent between big particles within an
OZ approach to the fluid structure, one must incorporate an accurate
theory for the bridge-diagrams.

The paper is laid out as follows: In Sec.\ \ref{sec:SM_pots} we describe
briefly our model fluid, the GCM, and the DFT used to calculate the solvent
density profiles and the SM potential between two big solute GCM particles.
Section\ \ref{sec:bridging} presents results for the density profiles
and SM potentials in the regime where there are thick adsorbed films around a
single big particle resulting in a bridging transition when two big
particles are sufficiently close together. In Sec.\ \ref{sec:bridging_th} we
present a simple analytic `capillarity' approximation which describes
qualitatively the onset of the bridging transitions that we find.
Section\ \ref{sec:2part} describes the effect of the formation of a
thick adsorbed film around a {\em pair} of big particles, in the portion of the
phase diagram where there is no thick film around a single big particle and
Sec.\ \ref{sec:bridging=bridge} describes our demonstration that bridging
between big particles is related to the bridge-function. Finally, in Sec.\
\ref{sec:conc}, we discuss our results and draw some conclusions.

\section{Model fluid and SM potentials}
\label{sec:SM_pots}
\begin{figure}
\noindent
\epsfig{figure=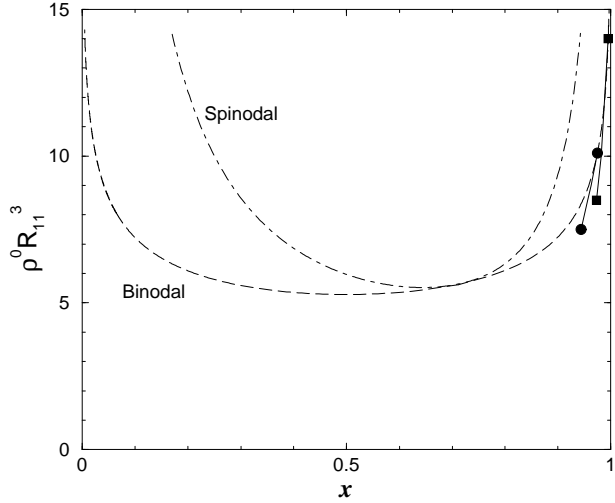,width=8cm}
\caption{The bulk phase diagram for a binary mixture of GCM particles with
$\epsilon_{12}/ \epsilon_{11} =0.944$ and $R_{22}/R_{11}=0.665$,
equivalent to a mixture of two polymers with length ratio 2:1 (see also 
Ref.~\onlinecite{Archer1}). $\rho^0$ is the
total density and $x$ is the concentration of the smaller species 2.
The solid line whose ends
are denoted by filled circles is the thin-thick adsorbed film
transition of the binary fluid adsorbed around a {\em single} big GCM particle
with pair potential parameters
$\beta \epsilon_{B1}=1.0$, $\beta \epsilon_{B2}=0.8$, $R_{B1}/R_{11}=5.0$ and
$R_{B2}/R_{11}=4.972$ -- see Ref.~\onlinecite{Archer5}.
It meets the binodal at the `wetting point' (upper circle)
with $x=0.975$ and  $\rho^0 R_{11}^3=10.1$ (note these values differ slightly
from the result quoted in Ref.~\onlinecite{Archer5} -- see footnote \cite{footnote})
and terminates at a critical point (lower circle) with $x=0.94$ and $\rho^0
R_{11}^3=7.5$. The solid line whose ends
are denoted by filled squares is the thin-thick adsorbed film
transition of the binary fluid adsorbed around a composite {\em pair} of the
same big GCM particles at zero separation $h=0$.
This transition line meets the binodal (upper square) at $x=0.995$ and  $\rho^0
R_{11}^3=14$ and terminates at a critical point
(lower square) with $x=0.973$ and $\rho^0 R_{11}^3=8.5$.}
\label{fig:phase_diag}
\end{figure}

We determine the SM potential between two big ($B$) Gaussian particles immersed
in a binary solvent of smaller Gaussian
particles. The GCM, in which the particles of species $i$
and $j$ interact via purely repulsive Gaussian potentials
\begin{equation}
v_{ij}(r) \, =\, \epsilon_{ij}\exp(-r^2/R_{ij}^2),
\label{eq:pair_pot}
\end{equation}
is a simple model for polymers in
solution \cite{Likos,LangJPCM,paper1,Archer1,Archer2,Dautenhahn} (in
particular, Ref.~\onlinecite{Likos} provides a good general introduction to
the GCM). For the binary GCM solvent we
choose pair potential parameters corresponding to a binary mixture of polymers
of length ratio 2:1, as were used in previous work on this model
fluid. \cite{Archer1,Archer2,Archer3,Archer5} The values are
$R_{22}/R_{11}=0.665$, $R_{12}/R_{11}=0.849$,
$\beta \epsilon_{11}=\beta \epsilon_{22}=2.0$ ($\beta =1/k_BT$) and
$\epsilon_{12}/\epsilon_{11}=0.944$. $R_{11}$ is the basic length scale in the
system. For this choice of parameters the binary mixture exhibits fluid-fluid
phase separation. The phase diagram of this binary solvent
is plotted in the total density, $\rho^0=\rho_1^0+\rho_2^0$, versus
concentration, $x=\rho_2^0/\rho^0$, plane in Fig.\ \ref{fig:phase_diag}
($\rho_{\nu}^0$ are the bulk densities of the small particles of species
$\nu=1,2$) -- see also Ref.~\onlinecite{Archer1}.

The SM potential between two big particles, labelled $A$ and $B$, with centres
at $\rr_A$ and $\rr_B$,
separated by a distance $h$, is given by the difference in the grand potential:
\begin{equation}
W_{AB}(h) \, =\, \Omega(| \rr_A-\rr_B| = h) \, - \, \Omega( |\rr_A-\rr_B| =
\infty).
\label{eq:SM_pot}
\end{equation}
This result can be re-expressed (trivially) in terms of excess grand potentials,
$\omega_{ex}^i \equiv \Omega-\Omega_b$, where $i=A,B$ and $\Omega_b$
is the grand potential of the bulk
solvent in the situation where there are no big particles present. Then,
\begin{equation}
W_{AB}(h) \, =\, \omega_{ex}^{AB}(| \rr_A-\rr_B| = h)
\, - \, \omega_{ex}^A\, - \, \omega_{ex}^B.
\label{eq:SM_pot_1}
\end{equation}
Note that $\omega_{ex}^i$, the excess grand potential for inserting
a single big particle of species $i$, is equal to
$\mu_{ex}^i$, the excess chemical potential of big species $i$ in
the limit of the bulk density of this species $\rho_i^0 \rightarrow 0$.
\cite{Archer3,Archer5,Roland} The effective pair
potential between two identical big particles is then the sum of the bare
interaction $v_{BB}(r)$ and the SM potential: 
\begin{equation}
v_{BB}^{eff}(h) \,=\, v_{BB}(h) \,+\, W_{BB}(h).
\label{eq:v_eff}
\end{equation}
Recall also, that
\begin{equation}
v_{BB}^{eff}(h)\,=\,-k_BT \ln g_{BB}(h),
\label{eq:log_g}
\end{equation}
where $g_{BB}$ is the big-big radial distribution function in the limit of the
big particle bulk density $\rho_B^0 \rightarrow 0$.
In the present work we use DFT to obtain the quantities $\omega_{ex}^{AB}(|
\rr_A-\rr_B| = h)$ and $\omega_{ex}^i$.

In DFT one calculates the solvent one body density profiles, $\{ \rho_{\nu}(\rr)
\}$, for a given set of external potentials, $\{V_{\nu}(\rr) \}$, by minimising
the grand potential functional: \cite{Bob}
\begin{eqnarray}
\Omega_V[\{\rho_{\nu}\}] \,& =&\,{\cal F}_{id}[\{\rho_{\nu}\}] \,+\, 
{\cal F}_{ex}[\{\rho_{\nu}\}]
-\sum_{\nu} \int\dr \, \rho_{\nu}(\rr) \, [\mu_{\nu}-V_{\nu}(\rr)],
\label{eq:grandpot}
\end{eqnarray}
where $\mu_{\nu}$ are the chemical potentials for the two species, $\nu=1,2$, of
solvent particles. The ideal gas part of the
intrinsic Helmholtz free energy functional is
\begin{equation}
{\cal F}_{id}[\{ \rho_{\nu} \}] \, = \, k_B T \sum_{\nu} \int\dr\,\,
\rho_{\nu}(\rr)\,[\ln (\Lambda_{\nu}^3 \rho_{\nu}(\rr))-1],
\label{eq:F_id}
\end{equation}
where $\Lambda_{\nu}$ is the thermal de Broglie wavelength of species $\nu$,
and ${\cal F}_{ex}[\{\rho_{\nu}\}]$ is the excess part of the
intrinsic Helmholtz free energy functional. Minimising (\ref{eq:grandpot})
together with (\ref{eq:F_id}) one
obtains the Euler-Lagrange equation
\begin{equation}
0 \,=\, k_BT \ln \Lambda^3_{\nu} \rho_{\nu}(\rr)
\,-\,k_BT c^{(1)}_{\nu}(\rr)
\,-\,\mu_{\nu}\,+\, V_{\nu}(\rr),
\label{eq:EL_eq}
\end{equation}
where
\begin{equation}
c^{(1)}_{\nu}(\rr)\,=\,-\beta \frac{\delta{\cal F}_{ex}[\{\rho_{\nu}\}]}
{\delta \rho_{\nu}(\rr)},
\label{eq:c1}
\end{equation}
is the one body direct correlation function, which is a functional of
$\{ \rho_{\nu} \}$. In an exact treatment the density profiles $\{ \rho_{\nu}
\}$ satisfying
(\ref{eq:EL_eq}) would yield the exact grand potential $\Omega$ as the minimum
of $\Omega_V$.
\cite{Bob} At this point we also recall that the two-body direct correlation
functions are given by the second functional derivative \cite{Bob}
\begin{equation}
c^{(2)}_{\nu,\xi}(\rr,\rr')\,=\,-\beta
\frac{\delta ^2{\cal F}_{ex}[\{\rho_{\nu}\}]}
{\delta \rho_{\nu}(\rr)\delta \rho_{\xi}(\rr')}.
\label{eq:c2}
\end{equation}

For the GCM the following approximate excess Helmholtz
free energy functional turns out, despite its
simplicity, to be remarkably accurate at high densities
$\rho^0 R_{11}^3 \gtrsim 5$:
\cite{Likos,LangJPCM,paper1,Archer1,Archer5,Archer6}
\begin{equation}
{\cal F}_{ex}^{RPA}[\{\rho_i\}] \,=\,
\frac{1}{2} \sum_{\nu,\xi} \int\dr\int\dr'\,\, 
\rho_{\nu}(\rr)\,\rho_{\xi}(\rr')
v_{\nu,\xi}(|\rr-\rr'|),
\label{eq:F_GCM}
\end{equation}
where $v_{\nu,\xi}(r)$ is the pair potential between the small solvent
particles of species $\nu$ and
$\xi$, given by Eq.\ (\ref{eq:pair_pot}).
The functional, Eq.\ (\ref{eq:F_GCM}), is that which generates the RPA
closure: $c^{(2),RPA}_{\nu,\xi}(\rr,\rr')=-\beta v_{\nu,\xi}(|\rr-\rr'|)$,
for the pair direct correlation functions. \cite{Likos,LangJPCM,paper1,Archer1}
The higher the density, the more accurate is the RPA for this soft core model.
\cite{Likos}

In the present work we choose the external potential to correspond to two fixed
big Gaussian particles of the same size, separated by a distance $h$:
\begin{eqnarray}
V_{\nu}(\rr) \,=& \epsilon_{B \nu} \exp(-(\rr+\mathbf{h}/2)^2/R_{B \nu}^2)
+ \epsilon_{B \nu} \exp(-(\rr-\mathbf{h}/2)^2/R_{B \nu}^2),
\label{eq:ext_pot}
\end{eqnarray}
with $\nu=1,2$ and
where $\mathbf{h}$ is a vector along the $z$-axis, with $|\mathbf{h}|=h$, i.e.\
the centres of the big particles are at $z= \pm h/2$.
Throughout the present study we choose the external potential parameters to be
$\beta \epsilon_{B1}=1.0$, $\beta \epsilon_{B2}=0.8$, $R_{B1}/R_{11}=5.0$ and
$R_{B2}/R_{11}=4.972$, the same values
as those used for the big-small particle pair
potentials in much of the work in Refs.~\onlinecite{Archer3,Archer5}. With this
external potential the solvent density profiles have cylindrical symmetry -- 
i.e.\ the density profiles are functions $\rho_{\nu}(z,r)$, where the $z$-axis 
runs through the centres of the two big particles, and $r$ is the radial
distance from the $z$-axis. If the external potential on the solvent were
exerted by hard particles, special care would be required to ensure that the
hard boundary is compatible with the grid of the numerical calculations in
order to avoid numerical artefacts in the contact
density.\cite{GrunbergKleinJCP1999} One would have to employ either matching
coordinate systems, such as the bispherical one used, e.g., in
Refs.~\onlinecite{Stark} and \onlinecite{Schlesener}, or  even more
sophisticated finite-element methods with adaptive
mesh-size.\cite{AndrienkoetalJCP2004}   
One of the appealing features of the {\em soft core} GCM used in the present
investigation is that we can avoid this problem and perform our calculations
on a cartesian grid in cylindrical coordinates.

\begin{figure}
\noindent
\epsfig{figure=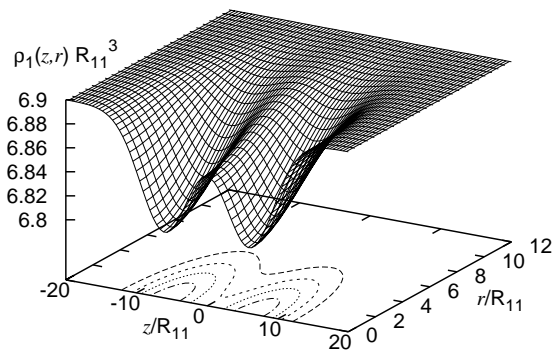,width=8.5cm}
\caption{The density profile of a one component fluid of Gaussian particles,
with bulk density $\rho_1^0 R_{11}^3 = 6.9$, around a pair of big Gaussian
particles, whose centres are located on the $z$-axis a
distance $h/R_{11}=12$ apart. The contours in the $z-r$ plane correspond to
$\rho_1(z,r)=6.82$, 6.84, 6.86 and 6.88.}
\label{fig:rho_one_comp}
\end{figure}

\begin{figure}
\noindent
\epsfig{figure=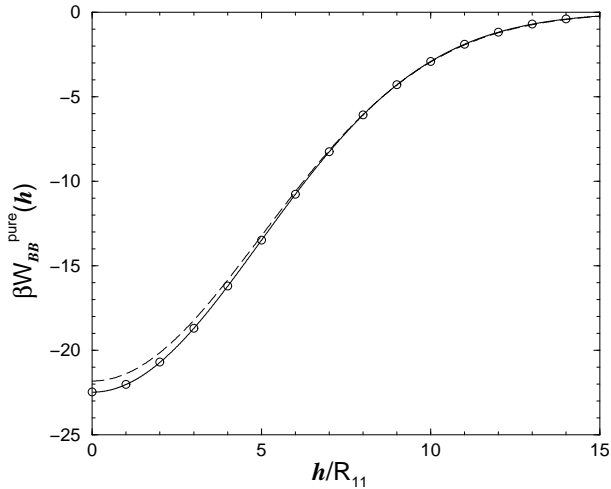,width=8cm}
\caption{The SM potential between two big GCM particles in a one component
solvent of small GCM particles, with bulk density $\rho_1^0 R_{11}^3 = 6.9$.
$h$ is the separation between the two big particles. The solid line is the DFT
insertion method results, the open circles are the results from the present
`brute-force' calculation (the two are almost indistinguishable) and the dashed
line is the analytic result, Eq.\ (\ref{eq:SM_analytic}), obtained in 
Ref.~\onlinecite{Archer5}.}
\label{fig:DeltaOmega_one_comp}
\end{figure}

In Fig.\ \ref{fig:rho_one_comp} we display a
typical density profile for a one component solvent of particles of species 1,
with the external potential given by Eq.\ (\ref{eq:ext_pot}) with
$h/R_{11}=12$. 
Having calculated the solvent density profiles for a given separation $h$ of
big particles, we can insert these into Eq.\ (\ref{eq:grandpot}) to
calculate the (excess) grand potential and the SM potential $W_{BB}(h)$ from
Eq.\ (\ref{eq:SM_pot_1}). In Fig.\ \ref{fig:DeltaOmega_one_comp} we display the
SM potential between two big GCM particles, calculated for a one component
solvent with bulk density $\rho_1^0 R_{11}^3=6.9$, i.e.\ the state point
corresponding to the profiles in Fig.\ \ref{fig:rho_one_comp}. Figure\
\ref{fig:DeltaOmega_one_comp} should be compared with Fig.\ 2 of
Ref.~\onlinecite{Archer5}. The open circles are the results from the present
`brute-force'
calculation. The solid line is the result obtained using the insertion method,
where one calculates only the solvent density profiles around an islolated,
single big particle and then uses the general result \cite{Roland}
\begin{equation}
\beta W_{BB}(h) \,=\, c_B^{(1)}(h \rightarrow \infty; \rho_B^0 \rightarrow 0)
\,-\, c_B^{(1)}(h; \rho_B^0 \rightarrow 0),
\label{eq:insertion_method}
\end{equation}
i.e.\ one calculates the difference in the excess chemical potential between
inserting the second big particle a distance $h$ from the first and inserting
it at $h=\infty$. As emphasised in the Introduction,
Eq.\ (\ref{eq:insertion_method}) is formally exact when we know the exact free
energy functional for a mixture of big and small particles. Here
we use the same RPA functional (\ref{eq:F_GCM}) extended to include a third
species $B$, in order to find an approximate
$c_B^{(1)}$ in Eq.\ (\ref{eq:insertion_method}) -- see Ref.~\onlinecite{Archer5}
for more details. The results from the two different routes are almost
indistinguishable for this point in the
phase diagram, and generally for other state points where no thick adsorbed
(wetting) films are present around the big
particles. The dashed line in Fig.\ \ref{fig:DeltaOmega_one_comp} corresponds to
the analytic approximation for $W_{BB}(h)$ presented in Ref.~\onlinecite{Archer5}:
\begin{equation}
\beta W_{BB}^{pure}(h) \, =\, -(\pi/2)^{3/2} \beta \epsilon_{B1} \rho^* R_{B1}^3
\exp(-h^2/2R_{B1}^2),
\label{eq:SM_analytic}
\end{equation}
where $\rho^*=\rho_1^0 \beta \epsilon_{B1}/(1+\pi^{3/2}\beta \epsilon_{11}
R_{11}^3 \rho_1^0)$. The agreement between this approximation and the result of
the full numerical DFT calculations is remarkably good.

\section{The SM potential when there are thick adsorbed films: bridging}
\label{sec:bridging}
\begin{figure}
\noindent

\epsfig{figure=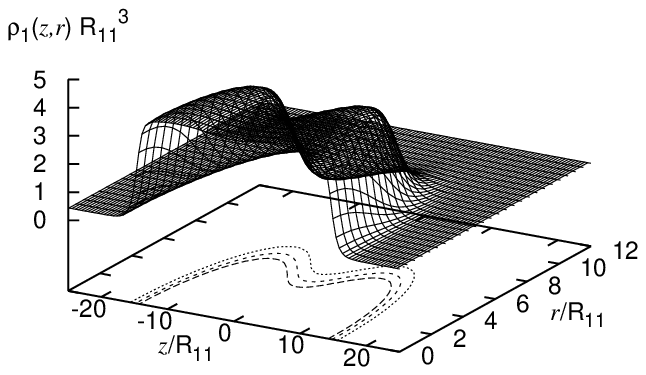,width=8.5cm}

\epsfig{figure=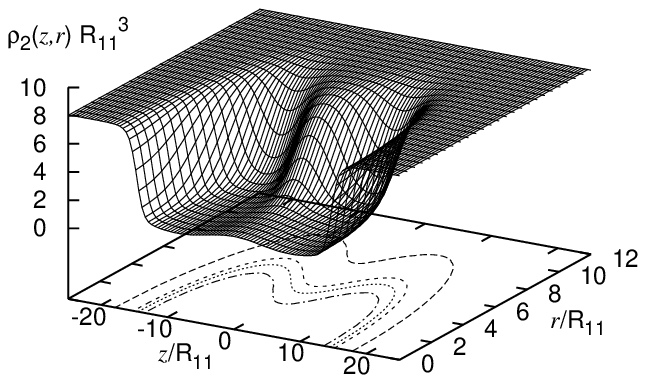,width=8.5cm}

\caption{Density profiles $\rho_{\nu}(z,r)$, $\nu=1,2$, for a solvent with total
density $\rho^0 R_{11}^3=8.5$ and concentration $x=0.948$, a state near to phase
separation located inside the single particle thin-thick adsorbed film
transition line (see Fig.\ \ref{fig:phase_diag}). The centres of the big
particles are a distance $h/R_{11}=17$ apart. Note the presence of thick
adsorbed (wetting) films and the
fluid bridge between the particles. The contours, plotted in the $z-r$ plane,
correspond to $\rho_1(z,r)R_{11}^3=1$, 2 and 3 and $\rho_2(z,r)R_{11}^3=2$ to 8
in increments of 2. The bridged configuration is the stable one for this value
of $h/R_{11}$.}
\label{fig:rhos_nr_bridging_heq16}
\end{figure}

\begin{figure}
\noindent

\epsfig{figure=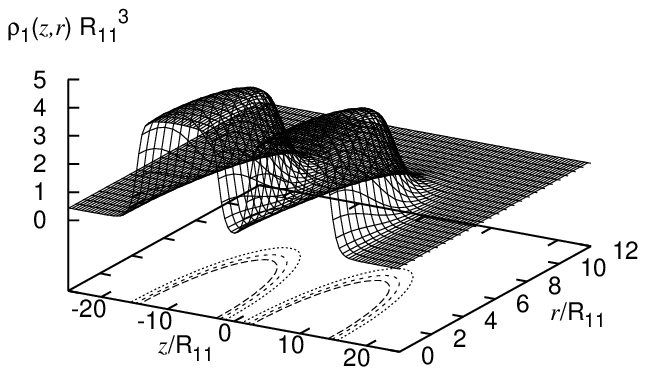,width=8.5cm}

\epsfig{figure=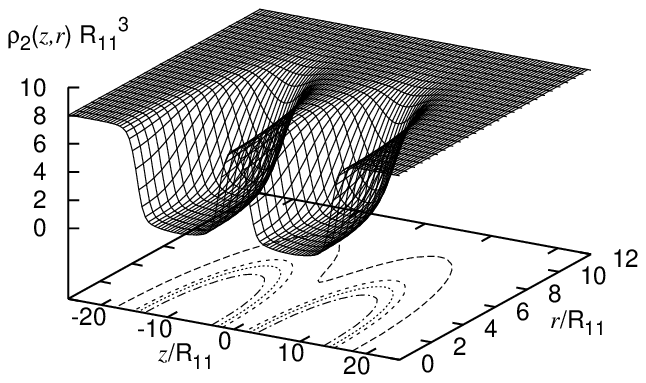,width=8.5cm}

\caption{Density profiles for the same state point and separation,
$h/R_{11}=17$, as Fig.\ \ref{fig:rhos_nr_bridging_heq16}, but now there is no
fluid bridge between the big particles. This configuration is metastable.}
\label{fig:rhos_nr_bridging_heq17}
\end{figure}

\begin{figure}
\noindent
\epsfig{figure=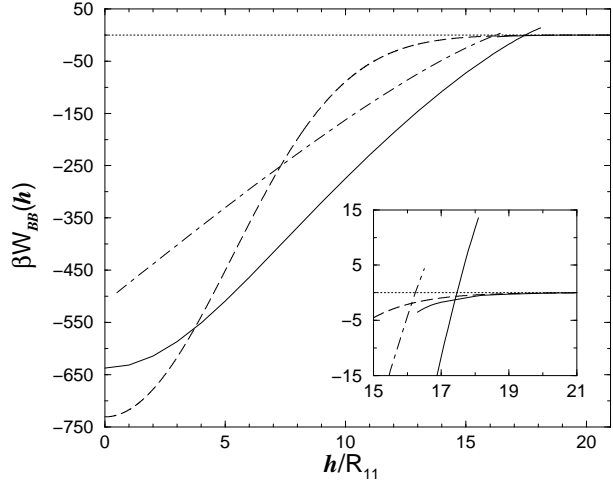,width=8cm}
\caption{The SM potential between two big GCM particles in a binary solvent
of smaller particles for the same state point as in Figs.\
\ref{fig:rhos_nr_bridging_heq16} and \ref{fig:rhos_nr_bridging_heq17},
i.e.\ with total bulk density $\rho^0
R_{11}^3 = 8.5$ and concentration $x=0.948$.
$h$ is the separation between the centres of the two big particles.
The dashed line is the result for $W_{BB}(h)$ obtained
using the insertion method, the
dot-dashed line is the `sharp-kink' result (see text, Sec.\
\ref{sec:bridging_th}) and the solid lines denote the results from the
`brute-force'
calculation. In the brute-force calculation, one finds that there are two
branches for $W_{BB}(h)$ (see inset for more detail), each with a metastable
portion. The branch with the smaller value of $W_{BB}(h)$ is stable. This
corresponds to the configuration with no bridge for $h>h_t$, and to the bridged
configuration for $h<h_t$. At $h_t/R_{11} = 17.4$, where the two branches cross,
there is a discontinuity in the gradient of $W_{BB}(h)$, i.e.\ there is a jump
in the SM force at this separation.}
\label{fig:DeltaOmega_bridging_inside_t-t}
\end{figure}

\begin{figure}
\noindent
\epsfig{figure=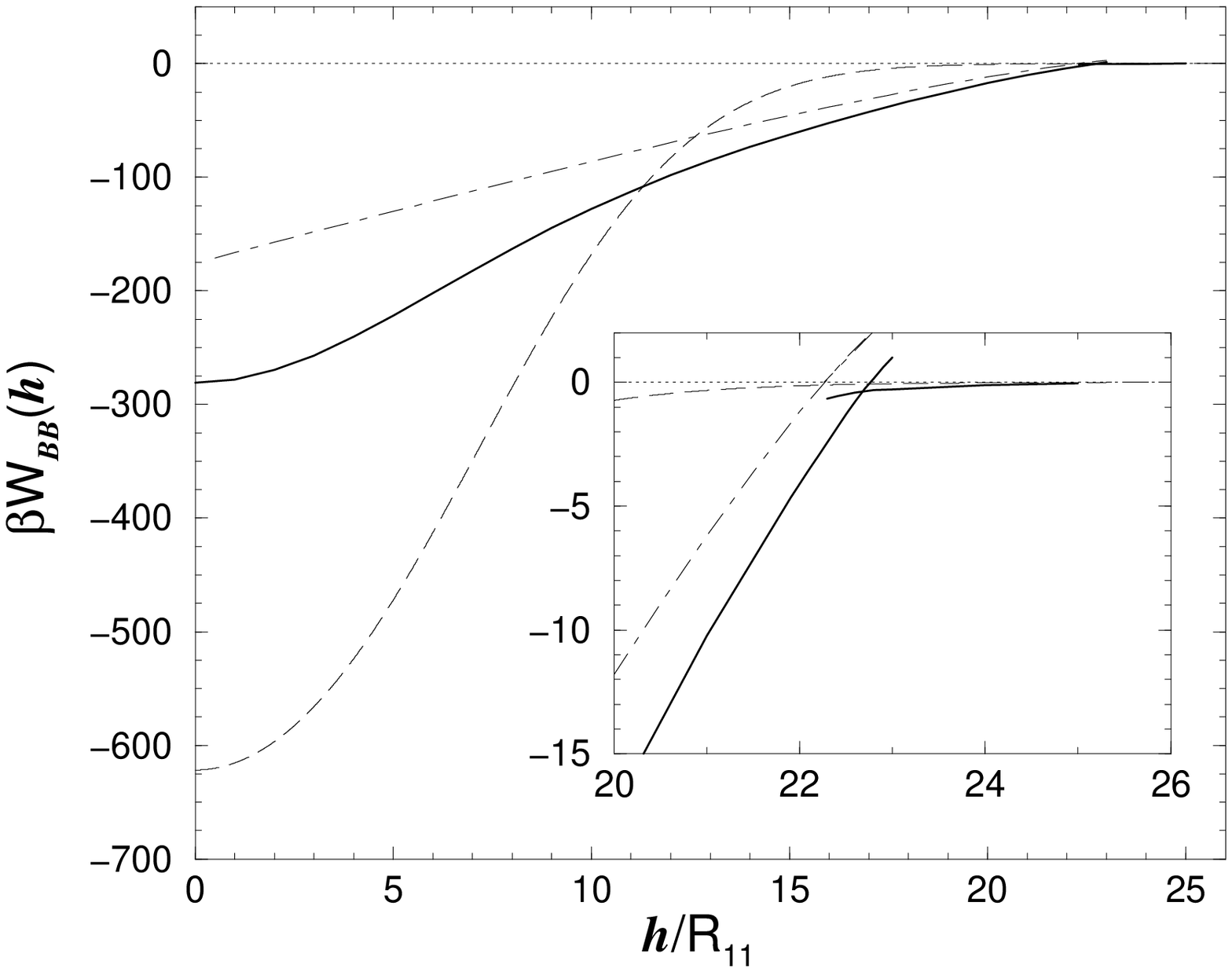,width=8cm}
\caption{The SM potential between two big GCM particles in a binary solvent
of smaller particles near to phase separation, with total bulk density $\rho^0
R_{11}^3 = 6.9$ and concentration $x=0.88$.
$h$ is the separation between the two big particles. The dashed line is the
result for the SM potential from the insertion method, the
solid lines are the results from the `brute-force'
calculation and the dot-dashed line is the `sharp-kink' result. In the
inset we display a magnification of $W_{BB}(h)$ for large $h$, showing the
two branches crossing at $h_t/R_{11}=22.7$ and giving rise to
a jump in the SM force.}
\label{fig:DeltaOmega_bridging}
\end{figure}

We now consider the case when thick adsorbed films develop around the big GCM
particles. The circumstances in which this can occur are discussed in 
Refs.~\onlinecite{Archer3,Archer5}. In general there can be thick adsorbed films when
the small solvent particles are in a state near
to phase separation. For the present mixture, the big GCM particles favour
species 1 of the small solvent particles, and so thick adsorbed films
of the coexisting phase rich in species 1 can develop when the big particles are
immersed in the solvent at a state point lying on
the right hand side of the binodal, which is poor in species 1.
In Refs.~\onlinecite{Archer3,Archer5} it was found that thick films develop via a
thin-thick transition out of bulk coexistence. The locus of these transitions is
shown as the solid line joining filled circles in Fig.\ \ref{fig:phase_diag}.
Note that this transition line meets the binodal at a `wetting point' whose 
density is somewhat higher than that quoted in Ref.~\onlinecite{Archer5}.
This discrepancy is associated with the existence of metastable minima in the
free energy. \cite{footnote} In Figs.\
\ref{fig:rhos_nr_bridging_heq16} and \ref{fig:rhos_nr_bridging_heq17} we
display density profiles calculated for a pair of big particles immersed in a
binary solvent of small GCM particles with bulk density $\rho^0 R_{11}^3=8.5$
and concentration $x=0.948$, a state point near to coexistence, located
inside the
single particle thin-thick adsorbed film transition line (see Fig.\
\ref{fig:phase_diag}). (Fig.\ 6 of Ref.~\onlinecite{Archer5} displays the solvent
density profiles around a {\em single} big particle for this state point.)
In Figs.\ \ref{fig:rhos_nr_bridging_heq16} and \ref{fig:rhos_nr_bridging_heq17} 
the centres of the big particles are a distance $h/R_{11}=17$
apart and there are thick adsorbed
wetting films around the big particles. However, in Fig.\
\ref{fig:rhos_nr_bridging_heq16} there is a fluid bridge between the two
particles whereas in Fig.\ \ref{fig:rhos_nr_bridging_heq17}, there is no fluid
bridge. This second set of profiles corresponds to a metastable situation. For
this state point the bridging transition occurs at a slightly larger separation
$h_t/R_{11}=17.4$; this is
where the bridged and unbridged configurations have equal grand potential.
In Fig.\ \ref{fig:DeltaOmega_bridging_inside_t-t} we display the SM potential
$W_{BB}(h)$ for this state point. There are two distinct branches, corresponding
to bridged and non-bridged configurations. For $h > h_t$ the unbridged
configuration is the stable one, whereas for $h < h_t$ the bridged configuration
becomes stable. Since the two branches of $W_{BB}(h)$ have different slopes
there is a discontinuity in the SM force, $-{\mathrm d}W_{BB}(h)/{\mathrm d}h$,
at $h_t$, the separation where the transition occurs. The extent of the
metastable portions is substantial; these extend well beyond the equilibrium
transition. This type of
metastability, with accompanying hysteresis, was also found by Stark \etal\
\cite{Stark} in their recent study of the bridging of the nematic wetting film
between two colloids immersed in the isotropic phase of a liquid crystal.
We display in Fig.\
\ref{fig:DeltaOmega_bridging} the SM potential calculated in the same way
for a different point in the phase diagram, closer to the solvent bulk critical
point, at a total density $\rho^0 R_{11}^3=6.9$ and concentration $x=0.88$.
This state point is also near to
bulk coexistence (see Fig.\ \ref{fig:phase_diag}). In both figures
\ref{fig:DeltaOmega_bridging_inside_t-t} and \ref{fig:DeltaOmega_bridging}
we compare the SM potential calculated using the present `brute-force' approach
(solid lines) with the results obtained using the
insertion method (dashed line) as described in Ref.~\onlinecite{Archer5}.
There is a significant
difference between the results from the two methods; the insertion
method does not capture the existence of two distinct branches of the grand
potential. Thus it does not appear to include explicitly the effects of a
bridging transition. The insertion method does predict very strongly
attractive SM
potentials, of a similar magnitude to those from full DFT, but does not yield
the correct shape or range for $W_{BB}(h)$. In contrast we recall from Sec.\
\ref{sec:SM_pots} that in the regime where there are no thick adsorbed films,
the results from the insertion method and the `brute-force' method are in good
agreement.

\section{Approximate model for the SM potential when bridging occurs}
\label{sec:bridging_th}
In Ref.~\onlinecite{Archer5} we found that when there was a thick adsorbed film
around a single big particle, a
good approximation for the excess grand potential of a {\em single} big GCM
particle immersed in a binary GCM solvent of small particles is:
\begin{equation}
\omega_{ex}^B \, \simeq \, \sum_{\nu=1}^2 \pi^{3/2} \epsilon_{B \nu} R_{B \nu}^3
\rho_{\nu}^{coex} \, + \, 4 \pi l^2 \gamma(l),
\label{eq:gamma_wett_film}
\end{equation}
where $\rho_{\nu}^{coex}$ are the solvent bulk densities in the coexisting
phase, i.e.\ the phase that forms the adsorbed film. $l$ is the thickness of
the adsorbed film ($l \sim R_{B \nu}$, but we determine its value by
calculating explicitly via DFT, the density profiles around a single
big particle) and
$\gamma(l)$ is the fluid-fluid surface tension, which we approximate by
$\gamma(\infty)$, the surface tension of the planar free interface
(this is calculated
using the approach presented in Ref.~\onlinecite{Archer1}). The first term in
Eq.\ (\ref{eq:gamma_wett_film}) is the excess grand potential for inserting a
single
big particle into the coexisting phase, obtained from the RPA bulk equation of
state, \cite{Archer5} and the second term is the
contribution from forming a spherical fluid-fluid interface. Generalising to
two big particles we might therefore expect the following approximation to
be reliable:
\begin{equation}
\omega_{ex}^{BB}(h) \, \simeq \, 2 \sum_{\nu=1}^2 \pi^{3/2} \epsilon_{B \nu}
R_{B \nu}^3 \rho_{\nu}^{coex} \, + \, A(l,h) \gamma'(l,h),
\label{eq:gamma_wett_film_2particles}
\end{equation}
where $A(l,h)$ is the surface area of the fluid-fluid interface between the
adsorbed film of the phase rich in species 1 which develops around the two big
particles and the bulk fluid rich in species 2. $\gamma'(l,h)$ is
the surface tension, which we again approximate by $\gamma(\infty)$, the planar
fluid-fluid interfacial tension. A similar sharp-kink or capillarity
approach was used in Ref.~\onlinecite{BauerPRE2000} to investigate
bridging for very big hard-core solute particles that induce thick
adsorbed (wetting) films but some new features
arise for soft core systems. At first sight we might expect
the first (single particle insertion) term in Eq.\
(\ref{eq:gamma_wett_film_2particles}) to be inaccurate
as $h \rightarrow 0$, when the big particles are strongly overlapping. However,
this is not the case. When
$h=0$ the first term in Eq.\ (\ref{eq:gamma_wett_film_2particles})
is accurate, since two big particles lying on top of each other result in an
external potential that has the same form as that due to a single big particle
with $\epsilon_{B \nu}$ twice the value
for one of the big particles taken alone.
In other words, if we take the first term in Eq.\ (\ref{eq:gamma_wett_film}) and
make the substitution $\epsilon_{B \nu} \rightarrow 2 \epsilon_{B \nu}$, then
we obtain the the same first term as in Eq.\
(\ref{eq:gamma_wett_film_2particles}). Given this observation the first term in
(\ref{eq:gamma_wett_film_2particles}) should be accurate for both large $h$ and
for $h=0$. Thus, by `continuity' we expect it to be accurate for all values of
$h$.
The overall accuracy of Eq.\ (\ref{eq:gamma_wett_film_2particles}) should
depend upon how accurately we determine the surface area $A(l,h)$ which appears
in the second term. 

Using equations
(\ref{eq:gamma_wett_film_2particles}), (\ref{eq:gamma_wett_film}) and
(\ref{eq:SM_pot_1}) we can obtain an expression for the SM potential:
\begin{equation}
W_{BB}(h) \, \simeq \, [A(l,h) \,-\, 8 \pi l^2] \gamma(\infty).
\label{eq:W_SK}
\end{equation}
We now present a simple model for $A(l,h)$, (see also
Ref.~\onlinecite{DzubiellaHansenJCP2004}) which we expect to be 
reliable for values of $h$ near to where the bridging transition occurs.

When $h \gg 2 l$, i.e.\ no fluid bridge is present, then $A(l,h) =
8 \pi l^2$, and Eq.\ (\ref{eq:W_SK}) gives $W_{BB}(h) =0$. When there is a
bridge we approximate the end sections of $A(l,h)$ by the surfaces of two
sections of spheres with radius $l$, and the bridge surface by the surface
generated by rotating the arc of a circle, of radius
$s$, about the axis passing through the centres of the end
sphere sections (the $z$-axis). We denote the width along the $z$-axis
of the bridge section by $2w$ and the diameter of the bridge section at the mid
point between the centres of the end sphere sections by $2d$. The surface area
of the two end spherical sections is $4 \pi l(l+h/2-w)$ and the surface area of
the bridge section is $4 \pi s (s+d) \arcsin(w/s)-4 \pi s w$. Requiring
continuity of the surfaces where the end and bridge sections meet and also
requiring continuity in the gradients at the point where these
sections join, we eliminate $s$ and $d$ to obtain the following expression
for the total surface area:
\begin{eqnarray}
A(l,h)&=& \,\frac{\displaystyle 2 \pi\, w\, l\, h}{\displaystyle (h/2-w)^2} \sqrt{ l^2-(h/2-w)^2}
\arcsin\left( \frac{\displaystyle h/2-w}{\displaystyle l} \right)\notag \\
 &&-\, \frac{\displaystyle 4 \pi\, w^2\, l}{\displaystyle h/2-w} \,
+\, 4 \pi\, l\, (l+h/2-w) \;.
\label{eq:area_eq}
\end{eqnarray}
We choose the value $w=w_0$ which minimises $A$, i.e.\
$\partial A/\partial w|_{w=w_0}=0$ 
and use this prescription for calculating
$A(l,h)$ with Eq.\ (\ref{eq:W_SK}) to calculate the SM potential
between two big GCM particles at state points
near to coexistence. For the case when the small
particle solvent has a total density $\rho^0 R_{11}^3 =8.5$ and concentration
$x=0.948$, corresponding to the full DFT calculation of the SM potential in
Fig.\ \ref{fig:DeltaOmega_bridging_inside_t-t}, we find that $\beta R_{11}^2
\gamma(\infty)=0.830$, and that  the film thickness $l/R_{11} \simeq 7$ (see
Fig.\ 6 in
Ref.~\onlinecite{Archer5}). Using these values in Eqs.\ (\ref{eq:area_eq}) and
(\ref{eq:W_SK}), we calculate the SM potential for this state point. The
result is the dot-dashed line displayed in Fig.\
\ref{fig:DeltaOmega_bridging_inside_t-t} which
is in good qualitative agreement with our results from the full `brute-force'
calculation of the SM potential, particularly for values of $h$ near to where
the bridging transition occurs. We also used this simple
approximation for the case when the small
particle solvent has a total density $\rho^0 R_{11}^3 =6.9$ and concentration
$x=0.88$, corresponding to the full DFT calculation of the SM potential in
Fig.\ \ref{fig:DeltaOmega_bridging}. For this state point
$\beta R_{11}^2 \gamma(\infty) = 0.152$, and $l/R_{11} =9.6$ (see Figs.\ 4 and
9 in Ref.~\onlinecite{Archer5}) and the SM potential is shown as
the dot-dashed line in Fig.\ \ref{fig:DeltaOmega_bridging}. Again, the results
are in qualitative agreement with those of the full calculation. In particular,
this simple approach provides a surprisingly accurate means of estimating the
value of $h$ at which the bridging transition will occur. If we assume that
bridging will only occur when $W_{BB}(h)<0$ the resulting values of $h_t$
underestimate the results of the full calculation by only a few percent in both
cases. Even for small values of $h$ the results of the sharp-kink approximation
for $W_{BB}(h)$
are of the correct magnitude. However, this approximation fails to reproduce the
correct shape of $W_{BB}(h)$ for small $h$.

\section{Thick adsorbed films on composite particles}
\label{sec:2part}
In the previous sections we considered only state points near the binodal
where we know that a single big GCM particle is `wet' by a thick adsorbed
film of the coexisting phase rich in species 1, i.e.\ state points inside or
below the single particle thin-thick adsorbed film transition line. However,
there can also be pronounced effects on the SM potential due
to the presence of thick adsorbed films for state points {\em outside}
the single particle thin-thick adsorbed film transition line (see Fig.\
\ref{fig:phase_diag}), where a single big particle immersed in the solvent
does not develop a thick adsorbed film. When two big particles are
sufficiently close together the resulting composite object can be sufficiently
large that a thick film is adsorbed. This effect is somewhat analogous to the
case for big hard-core solute particles, where for certain state points for
which
no thick adsorbed films are present, capillary condensation of the coexisting
phase can occur in the space between the two big particles, provided these come
sufficiently close together. \cite{BauerPRE2000,Stark}
We cannot strictly describe the phenomenon we observe as
capillary condensation because the big particles that we consider in the present
work have soft cores. Nevertheless, the present phenomenon
has a similar effect on the SM potential, i.e.\ there is a jump in
the SM force on reducing the separation $h$. As mentioned above, this phenomenon
occurs outside (but close to) the single particle thin-thick adsorbed film
transition line. However, its occurrence is restricted to a particular region of
the phase diagram. If one considers two big particles with
full overlap ($h=0$) one can calculate the thin-thick adsorbed film transition
line for this composite object. This line is higher in total density than the
corresponding single particle
transition line (see Fig.\ \ref{fig:phase_diag}) and
serves as an upper bound for the regime where `capillary condensation' occurs;
the latter is restricted to the region betwen the two transition lines.

The solvent density profiles around two big particles with $h$ sufficiently
small that this `condensation' has occurred are very similar in form to the
profiles in Fig.\ \ref{fig:rhos_nr_bridging_heq16}, i.e.\ the
`condensation' does not just occur in the space between the two big particles,
as would be the case with a pair of hard-core big particles.
Rather, due to the soft core nature of the GCM, the adsorbed film spreads around
the whole region in which the two big particles are situated.

\begin{figure}
\noindent
\epsfig{figure=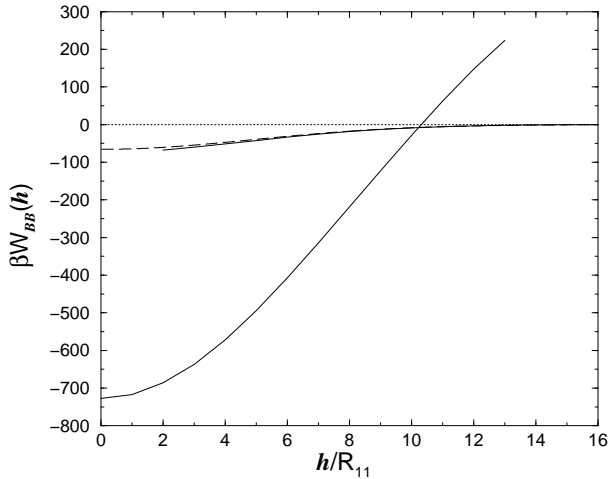,width=8cm}
\caption{The SM potential between two big GCM particles in a binary solvent
of smaller particles with total bulk density $\rho^0
R_{11}^3 = 11$ and concentration $x=0.983$ (this state point is at bulk
coexistence, outside the single particle thin-thick adsorbed film
transition line, but inside the thin-thick adsorbed film transition line for a
composite
pair of completely overlapping big particles -- see Fig.\ \ref{fig:phase_diag}).
$h$ is the separation between the centres of the two big GCM particles. The
dashed line is the result for $W_{BB}(h)$ obtained using the insertion method
and the solid lines denote the results from the `brute-force'
method. For this state point a single big particle does not develop thick
adsorbed film, but when two big particles are
sufficiently close together the resulting composite object can develop a thick
adsorbed film. The two branches of $W_{BB}(h)$ correspond to configurations
without adsorbed films (stable at large $h$) and with films (stable at small
$h$). These cross at $h/R_{11}=10.2$, resulting in a discontinuity in the
gradient of $W_{BB}(h)$ and a jump in the SM force.}
\label{fig:DeltaOmega_outside_t-t}
\end{figure}

In Fig.\ \ref{fig:DeltaOmega_outside_t-t} we display the
SM potential between two big GCM particles in a binary solvent
of smaller particles with total bulk density $\rho^0
R_{11}^3 = 11$ and concentration $x=0.983$. This state point is located at bulk
coexistence above the single particle thin-thick adsorbed film transition line
but inside the transition line for the composite particle -- see Fig.\
\ref{fig:phase_diag}. For large values of $h$ the
SM potential calculated via the `brute-force' approach is in good
agreement with the results from the insertion method. At this state point the
insertion method does not include any effects of thick adsorbed films since the
inputs into this theory are the density profiles around a single big particle;
for this state point a single big particle
has no thick adsorbed film. However, as $h$ is decreased the results of the
full DFT calculation show that there is a
discontinuity in the gradient of $W_{BB}(h)$ due to the formation of a thick
adsorbed film around the two particles. The change in the SM
potential is very pronounced; the potential becomes much more strongly
attractive -- see Fig.\ \ref{fig:DeltaOmega_outside_t-t}.
The insertion method (dashed line) accounts extremely well for the large $h$
behaviour of the SM potential. It also describes accurately the metastable
portion of $W_{BB}(h)$ for $h$ below the transition value. However, it fails
completely to describe the stable, strongly attractive portion arising from the
formation of the thick adsorbed film around the two particles; it underestimates
the strength of the attraction by a factor of about 10. This is not too
surprising given that this method inputs only the density profiles around a
single big particle and that these exhibit no thick adsorbed films for this
state point.

\section{Bridging and the bridge function}
\label{sec:bridging=bridge}

We recall that the SM potential $W_{BB}(r)$ 
is related via Eqs.~(\ref{eq:v_eff}) and (\ref{eq:log_g}) to the pair
correlation function between solute particles, $g_{BB}(r)$,
in a bulk ternary mixture which consists of a single, big solute species $B$ and
two solvent species, considered in the dilute limit
of solute, $\rho_B \to 0$. Since integral equations are a standard tool
to determine bulk pair correlation functions
in the theory of classical liquids, \cite{HM}
it is natural to analyse the SM potential within this framework.
However, we recall from the outset that whilst integral equation theories have
achieved remarkable precision in the description of one--component bulk fluids,
integral equation closure approximations are generally less reliable
in multi--component mixtures, especially for situations where the size of one
component becomes much larger than the others leading to the possibility of
thick film adsorption or wetting phenomena or, in the case of hard-sphere
mixtures, to depletion phenomena.

Before we analyse $g_{BB}(r)$ in the ternary mixture, it is instructive  to
point out some features of the two--component solvent mixture which provide a
relationship between the hypernetted chain (HNC) integral equations and the RPA
density functional used in the present work. Diagrammatic analysis yields the
following standard relationships between the pair correlation functions in a
homogeneous (bulk) mixture: \cite{HM}
\begin{eqnarray}
h_{ij}(r) -c_{ij}^{(2)}(r) &=& \sum_{k=1,2} \rho_k^0 \int \dr'
h_{ik}(|\rr-\rr'|)c_{kj}^{(2)}(r'),
\label{eq:OZ_eq} \\
  \ln g_{ij}(r) + \beta v_{ij}(r) &=& h_{ij}(r)-c_{ij}^{(2)}(r) +b_{ij}(r), \quad
\label{closure}
\end{eqnarray}
where $h_{ij}(r)=g_{ij}(r)-1$. The first equation is the OZ equation for binary
mixtures, and the second
provides the formally exact closure to the OZ equation in terms of the 
(generally unknown) bridge function $b_{ij}(r)$. 
The bulk densities of the two solvent species are denoted by $\rho_k^0$
($k=1,2$). It is a special feature of the 
binary GCM (or related soft--core models) 
that its pair correlation functions are very well described within the
HNC approximation
\cite{Likos,LangJPCM,paper1,Ard,Bolhuis,Archer1,Archer2,Finken,Archer6,Archer4}
which amounts to setting $b_{ij}(r)=0$. We
denote the corresponding solution for the pair direct correlation function by 
$c^{(2),HNC}_{ij}(r)$. The relation to density functional theory follows by
noting that
the HNC equations, Eq.~(\ref{closure}) with $b_{ij}(r)=0$, are identical to the
test particle equations obtained from a DFT with the
excess free energy functional \cite{Oettel}
\begin{eqnarray}
{\cal F}_{ex}^{HNC}[\{\rho_i\}] &=& A_{ex}(\{\rho_i^0\})
+ \sum_{i=1,2} \int \dr \mu_i^{HNC} \Delta \rho_i(\rr)
- \nonumber \\ 
 &&\frac{1}{2\beta} \int \dr \int \dr'
\sum_{ij=1,2} c^{(2),HNC}_{ij} (|\rr-\rr'|)
\Delta \rho_i(\rr) \Delta \rho_j(\rr'), \quad
\label{eq:fhnc} 
\end{eqnarray}
corresponding to a Taylor expansion to quadratic order in $\Delta \rho_i(\rr)$
about the bulk densities.
The test particle equations follow by choosing as external
potentials the interparticle potential $v_{ji}(r)$, minimising the HNC grand
potential functional with respect to $\rho_j(r)$ and identifying
$g_{ji}(r) \equiv \rho_j(r)/\rho_j^0$. 
In Eq.~(\ref{eq:fhnc}), $\Delta \rho_i(\rr)=\rho_i(\rr)-\rho_i^0$ and 
$A_{ex}(\{\rho_i^0\})$ denotes
the excess Helmholtz free energy of the bulk solvent. The HNC chemical potential
is given by
\begin{eqnarray}
 \beta \mu_i^{HNC} &=& \sum_{j=1,2} \rho_j^0
 \int d\rr \left( \frac{1}{2}h_{ji}(r)
 [h_{ji}(r)-c^{(2),HNC}_{ji}(r)] - c^{(2),HNC}_{ji}(r) \right)\;.
 \label{eq:mu_hnc}
\end{eqnarray}
Previous results for the GCM showed that the pair correlation functions obtained
from the HNC were similar to those obtained from the RPA
\cite{Likos,LangJPCM,paper1,Archer4} and that the fluid-fluid binodals from the
RPA and the HNC approximation were close to each other. \cite{Archer4}
If one neglects the weak density dependence of
$c^{(2),HNC}_{ji}(r)$ and sets $c^{(2),HNC}_{ji}(r) \simeq c^{(2),RPA}_{ji}(r)
=-\beta v_{ji}(r)$ then one can show for the binary mixture, $i=1,2$:
\begin{equation}
  {\cal F}_{ex}^{HNC}[\{\rho_i\}] \simeq {\cal F}_{ex}^{RPA}[\{\rho_i\}]\,
\end{equation}
where ${\cal F}_{ex}^{RPA}$ is the RPA functional defined in
Eq.~(\ref{eq:F_GCM}).

We have seen in earlier sections that for a fixed big Gaussian particle exerting
an external potential on the solvent 
close to coexistence, the RPA functional accounts for the formation of a thick
adsorbed film. It also accounts for complete wetting at a planar wall.
\cite{Archer2} Owing to the weak density dependence of $ c^{(2),HNC}_{ij}(r)$
we also expect the HNC functional to describe thick film formation and complete
wetting. (This is in sharp contrast to simple
fluids of the Lennard--Jones type
where the harshly repulsive core in the interatomic potential induces a strong
density dependence of the direct correlation function $c^{(2)}(r)$ and the HNC
functional (\ref{eq:fhnc}) fails to account for complete wetting
\cite{Oettel,EvansetalMolecP1983}). 

Explicit minimisation of the HNC functional for a binary mixture in the presence
of an external  potential due to a single solute particle yields the HNC
solute--solvent integral equations.
These can also be derived from the test particle equations 
of the HNC functional for the {\em ternary} mixture of binary solvent
plus solute {\em in the dilute limit} of the solute, $\rho_B \to 0$.
This functional is linear in $\rho_B(\rr)$ and is, at most, quadratic in the
other density profiles. It is given by
\begin{eqnarray}
{\cal F}_{ex,tern}^{HNC} &=&  {\cal F}_{ex}^{HNC} + \mu_B^{HNC}
\int \dr\, \rho_B(\rr) \nonumber \qquad  \\
&&- \frac{1}{2\beta} \int \dr \int \dr' \sum_{i=1,2}
c^{(2),HNC}_{iB} (|\rr-\rr'|)
\Delta \rho_i(\rr)\, \rho_B(\rr')\;. 
\label{eq:fhnc3}
\end{eqnarray}
Here, $\mu_B(\{\rho_i^0\})$ is the HNC insertion free energy (chemical
potential) for inserting a single solute particle into the bulk solvent with
densities $\rho_i^0$ ($i=1,2$). Analogously to Eq.~(\ref{eq:mu_hnc}),
$\mu_B^{HNC}$ is given by
\begin{eqnarray}
 \beta \mu_B^{HNC} &=& \sum_{i=1,2} \rho_i^0
 \int \dr \left( \frac{1}{2}h_{iB}(r)
 [h_{iB}(r)-c^{(2),HNC}_{iB}(r)] 
  - c^{(2),HNC}_{iB}(r) \right)\;, 
\end{eqnarray}
where the solute--solvent pair correlation function $h_{iB}(r)$ and the direct
correlation function $c^{(2),HNC}_{iB}(r)$ are determined 
by solving the solvent--solvent and solute--solvent HNC equations.
In the dilute limit of solute, the solute--solvent
direct correlation function satisfies the OZ equation  
\begin{equation}
c_{iB}^{(2)}(r)\,=\, h_{iB}(r) \,-\sum_{j=1,2} \rho_{j}^0 \int \dr'
h_{Bj }(|\rr-\rr'|)c_{ji}^{(2)}(r'),
 \label{eq:OZ_eq_2}
\end{equation}
for $i=1,2$. In this treatment thick adsorbed films can
develop around a big solute particle and this
is manifest in the density profiles of the two solvent species and thus in
$h_{Bj}(r)$. It follows from Eq.\ (\ref{eq:OZ_eq_2}) that
information about thick films is fed into $c_{iB}^{(2)}(r)$. We can deduce that 
whenever thick film formation occurs, $c_{iB}^{(2),HNC}(r)$ can be very
different from the RPA result $-\beta v_{iB}(r)$.

We turn attention now to the solute--solute correlation functions. These are
generated by employing ${\cal F}_{ex,tern}^{HNC}$, fixing $v_{BB}(r)$ as the
external potential and minimising the grand potential functional
with respect to $\rho_B(r)$. One finds
\begin{eqnarray} 
\ln g_{BB}(r) +\beta v_{BB}(r) 
= \sum_{i=1,2} \rho_i^0 \int \dr'
h_{Bi }(|\rr-\rr'|)c_{iB}^{(2),HNC}(r') \; . 
\label{eq:lng_BB}
\end{eqnarray}
If one now employs the mixture OZ equations in the limit $\rho_B \to 0$ one
obtains
\begin{equation} 
  \ln g_{BB}(r) +\beta v_{BB}(r) = h_{BB}(r) - c^{(2)}_{BB}(r) \; . 
\label{eq:lng_BB_2}
\end{equation}
Note that the right hand side of Eq.\ (\ref{eq:lng_BB}) depends on the
solute--solvent correlation functions
$h_{Bi}(r)$ and and $c_{iB}^{(2),HNC}(r)$. The
former quantity is, essentially, the density profile of species $i$ around a
single big particle determined by minimising the HNC functional and the latter
is given by the OZ equation (\ref{eq:OZ_eq_2}). One might expect both quantities
to be given accurately by the HNC treatment. The $g_{BB}(r)$ resulting from Eq.\
(\ref{eq:lng_BB}) yields, via Eqs.\ (\ref{eq:v_eff}) and (\ref{eq:log_g}), an
SM potential which we refer to as $W_{BB}^{HNC}(r)$ since this is consistent
with the fact that $g_{BB}(r)$ satisfies Eq.\ (\ref{eq:lng_BB_2}), the HNC
equation for big-big correlations; the latter sets the bridge function
$b_{BB}(r)=0$.

As the HNC inputs only the pair direct correlation
functions $c_{ij}^{(2),HNC}(r)$
of the {\em small} solvent species, which should be well described by their RPA
counterparts, we adopt the following procedure: determine the density profiles
of the two small species around a single big particle by minimising the RPA
grand potential functional, Eqs.\ (\ref{eq:grandpot}) and (\ref{eq:F_GCM}), and
use these as input for $h_{Bi}(r)$, along with
$c_{ij}^{(2),RPA}(r)$ for the solvent--solvent direct correlation functions, in
Eq.\ (\ref{eq:OZ_eq_2}). The resulting $c_{iB}^{(2)}(r)$ are then used in Eq.\
(\ref{eq:lng_BB}) to calculate $g_{BB}(r)$ and, hence, the SM potential -- which
should be very close to $W_{BB}^{HNC}(r)$.
We find that $W_{BB}^{HNC}(r)$, for large $r$, is almost identical to the
branch of $W_{BB}(r)$ obtained using the `brute-force' DFT method presented in
Secs.\ \ref{sec:SM_pots} and \ref{sec:bridging}, for which there is {\em no
fluid bridge}. In other words, when there is no fluid
bridge, i.e.\ for $r \equiv h>h_t$, where $h_t=h_t(\rho_1^0,\rho_2^0)$ is the
separation at which the bridging transition occurs,
$W_{BB}^{HNC}(r) \simeq W_{BB}(r)$ and we can infer that the HNC approximation
$b_{BB}(r)\simeq 0$ is valid. However, for $h<h_t$ we find $W_{BB}^{HNC}(r)$ is
very different from $W_{BB}(r)$, indicating that the bridge function
$b_{BB}(r)$, omitted from this analysis, must be
substantial for $h<h_t$. Thus we have demonstrated that $b_{BB}(r)$ must play a
significant role in determining the fluid structure when there is bridging.

We conclude that the ternary HNC functional, Eq.~(\ref{eq:fhnc3}), describes
correctly the bulk solvent--solvent correlations
and captures thick film formation in the solute--solvent correlations
with a vanishing solute--solvent bridge function, $b_{iB}(r)=0$. 
For the solute--solute correlations when bridging is not present
the HNC assumption $b_{BB}(r)=0$ remains accurate but this approximation fails
completely when bridging is present.
This means that a more sophisticated theory should include in the ternary
functional terms 
proportional to $\rho_B(\rr)\,\Delta \rho_i(\rr')\,\Delta \rho_j(\bf r'')$ and
higher orders. These will become important near the onset of the transition. 


\section{Discussion and conclusions}
\label{sec:conc}

Using `brute-force' DFT we have calculated the SM potential $W_{BB}(h)$
between a pair of big GCM particles in a binary solvent of smaller GCM
particles. In particular, we have focused on the regime where
the big particles are immersed in the binary solvent near to bulk phase
separation,
where thick films of the coexisting solvent phase adsorbed around the big
particles influence strongly the SM potential. It is in this regime that we
find bridging transitions. We show that the insertion method for calculating the
SM potential used in Refs.~\onlinecite{Archer3,Archer5}, and which is based on the
ternary version of the RPA functional (\ref{eq:F_GCM}), is unable to incorporate
the effects of bridging. This method does provide an accurate approximation for
$W_{BB}(h)$ for solvent state points away from the binodal. The bridging that we
find is of two types: i) that due to
the joining of thick adsorbed films around the individual big particles,
described in Sec.\ \ref{sec:bridging}, and ii) that due to local condensation
around a pair of particles, described in Sec.\ \ref{sec:2part}. Both result in a
change in slope of $W_{BB}(h)$ at a separation $h=h_t$ and therefore a jump in
the SM force at $h=h_t$.

Within our mean-field theory, bridging manifests itself
as a sharp (first-order) transition. However,
this cannot be the case in reality since the bridging transition involves a
finite number of particles and therefore fluctuation effects will
round the transition (see discussion in Ref.~\onlinecite{BauerPRE2000}).
We can make a crude estimate of the extent of rounding effects by arguing that
fluctuations should only be relevant when $|W_{BB}^{{\mathrm b}{\mathrm r.}}(h)
- W_{BB}^{{\mathrm n}{\mathrm o} \, \, {\mathrm b}{\mathrm r.}}(h)| \lesssim
k_BT$, where $W_{BB}^{{\mathrm b}{\mathrm r.}}(h)$ denotes the branch of
$W_{BB}(h)$ where there is a fluid bridge and $W_{BB}^{{\mathrm n}{\mathrm o} \,
\, {\mathrm b}{\mathrm r.}}(h)$ the branch without a fluid
bridge. From this inequality we can obtain the width,
$\delta h_t$, over which the transition at $h_t$ will be smeared. We find that
$\delta h_t/h_t \sim 10^{-2}$ for the state points corresponding to the SM
potentials displayed in Figs.\ \ref{fig:DeltaOmega_bridging_inside_t-t}
and \ref{fig:DeltaOmega_bridging}. This measure of the rounding becomes smaller
for solvent state points further removed from the bulk critical point. For
bigger solute particles we also expect the extent of the rounding to become
smaller. At first sight our estimate of the rounding may
seem surprisingly small, bearing in mind that the size ratio between the big
solute and
small solvent particles is only about 7:1. However, due to the soft-core nature
of the GCM fluid, the solvent density is high and the number of particles
involved in the bridging transition can be large. This demonstrates one of
the important differences between the soft-core GCM and more typical hard-core
fluid systems: For hard-core particles one would not find thick adsorbed
films of the solvent were the size ratio between the solute and solvent only
7:1. Typically the solute must be of order 50 or more times larger than the
solvent particles for wetting phenomena to become significant -- see
also the discussion in Ref.~\onlinecite{Archer5}.

Our analysis in Sec.\ \ref{sec:bridging=bridge} demonstrates that in order to
incorporate bridging into a full (ternary) mixture theory, one must implement an
accurate theory for the fluid bridge functions; in particular for
the solute-solvent and solute-solute bridge functions $b_{Bi}(r)$ and
$b_{BB}(r)$. $b_{BB}(r)$, the solute-solute bridge
function, remains little understood but must play a crucial role when there are
thick adsorbed films surrounding the big particles.
That the bridge functions are required
highlights the essential many-body nature of the effective interaction between
the big solute particles. Hence, it is not surprising that
the insertion method combined with the ternary version of the RPA
functional (\ref{eq:F_GCM}) is unable to incorporate the effects of bridging on
the SM potential. We re-iterate that the insertion method is formally exact; it
is its use with an approximate functional which leads to neglect of the key
features of bridging. In order to obtain
insight as to what is required in a theory for the full mixture
Helmholtz free energy functional which incorporates the effect of bridging,
we consider the exact inhomogeneous Kirkwood-Hill formula
\cite{HendersonMolecPhys83,Archer5} (recall Eq.\ (\ref{eq:c1})):
\begin{eqnarray}
c_B^{(1)}(\rr) = 
-\sum_{\nu=1}^2 \int_0^1 {\rm d}\lambda \int
\dr'\rho_{\nu}(\rr') g_{B \nu}(\rr,\rr';\lambda) \beta v_{B \nu}(|\rr-\rr'|),
\label{eq:Kirkwood_inhom}
\end{eqnarray}
for the one-body direct correlation function of the big solute particles in the
limit $\rho_B \rightarrow 0$. $v_{B
\nu}(r)$ are the big-small pair potentials and the parameter $\lambda$, with
$0 \leq \lambda \leq 1$, is used to `turn on' the effect of the inserted
big particle via
the potential $\lambda v_{B \nu}(r)$. One calculates the solvent response
through the inhomogeneous big-small pair distribution function
$g_{B \nu}(\rr,\rr';\lambda)$, as $\lambda$ is increased from 0 to 1.
Combining Eq.\ (\ref{eq:Kirkwood_inhom}) with Eq.\
(\ref{eq:insertion_method}) one obtains an exact expression for $W_{BB}(r)$,
given by Eq.\ (70) of Ref.~\onlinecite{Archer5}.
Consider the case when the solvent is near coexistence at a
state point below the single big particle thin-thick adsorbed film transition
line. If one calculates $W_{BB}(h)$ via Eq.\ (\ref{eq:Kirkwood_inhom}), then
$g_{B \nu}(\rr,\rr';\lambda=0)$ will correspond to the distribution arising from
a fixed single big particle located at
$\rr=-\mathbf{h}/2$ exerting an external potential on the solvent.
This big particle will be surrounded by a thick adsorbed film. Then,
`turning on' the effect of the second big particle (by increasing $\lambda$ from
zero) located at $\rr=+\mathbf{h}/2$ one could perhaps envisage the
situation where there might
be two `jumps' in $g_{B \nu}(\rr,\rr';\lambda)$ for a particular value of
$h=|\mathbf{h}|$. The first would be at $\lambda=\lambda_1$, when
the potential $\lambda_1 v_{B \nu}(r)$ becomes sufficiently strong
to induce condensation of the coexisting solvent phase around
this second big particle. This `jump' in $g_{B \nu}(\rr,\rr';\lambda)$ could
then be followed by a second jump at $\lambda=\lambda_2$
($\lambda_1<\lambda_2<1$), when a fluid bridge forms between the two big
particles. That such complex phenomena must be described by $c^{(1)}_B(\rr)$,
which is obtained by taking one functional derivative of the excess Helmholtz
free energy functional Eq.\ (\ref{eq:c1}), attests to the degree of
sophistication required in the ternary mixture functional
${\cal F}_{ex}[\{\rho_i\}]$ if this is to incorporate bridging. By employing
the RPA functional in the insertion approach one incorporates the effect of
the thick film surrounding the first big particle via 
$\rho_{\nu}(\rr)$ in Eq.\ (\ref{eq:Kirkwood_inhom}), but neglects
the effect of the thick film around the second big particle by setting
$g_{B \nu}(\rr,\rr';\lambda)=1$ for all $\lambda$. \cite{Archer5} One is
thereby unable to incorporate the effect of
bridging on the SM potential. This point is highlighted further by the case
described in Sec.\ \ref{sec:2part}, where a single big particle has no thick
adsorbed film and the bridging arises from condensation
around a {\em pair} of big particles. In this situation
all the information about bridging/wetting must be generated in
$c_B^{(1)}(\rr)$ from a source other than the solvent density profiles around a
single big particle, i.e.\ from subtle correlations in the inhomogeneous
solvent. Incorporating such correlations is a tall order for a theory!

The simple capillarity (or sharp-kink) approximation used in
Sec.\ \ref{sec:bridging_th} to
provide an approximate theory for when bridging occurs between two thick
adsorbed films seems to be quite good.
The simple form, Eq.\ (\ref{eq:W_SK}), taken with Eq.\ (\ref{eq:area_eq}), is
surprisingly reliable in determining an approximation for $h_t$, the
separation between the big particles at which bridging occurs, as well as
providing a reasonable approximation for the slope of $W_{BB}(h)$ near the onset
of the
bridged configuration, i.e.\ it provides quite a good approximation for the SM
force at $h \sim h_t$. The capillarity approximation is not reliable
for small $h$.
Here the shape of the SM potential determined from the capillarity approximation
is completely wrong, and therefore the SM force obtained from this approximation
will be completely unreliable -- see Figs.\
\ref{fig:DeltaOmega_bridging_inside_t-t} and \ref{fig:DeltaOmega_bridging}. The
`brute-force' calculation shows that as $h \rightarrow 0$, the SM force
$\rightarrow 0$, whereas the capillarity approximation shows the SM force
tending to a non-zero constant value as $h \rightarrow 0$. We believe the origin
of this failure lies in our simple approximation (\ref{eq:area_eq}) for the area
of the fluid-fluid interface.

One issue we have not raised so far is what does one
take for $v_{BB}(r)$, the bare big-big pair potential? This does not enter our
calculation of the SM potential, since it is only $v_{B \nu}(r)$, the
big-small pair potentials, that are involved; the big
particles are treated as external potentials. Therefore, in principle,
$v_{BB}(r)$ could take any form, although choosing a bare
potential with a hard core would
be inconsistent with the soft-core nature of $v_{B \nu}(r)$. A Gaussian
potential of the form given in Eq.\ (\ref{eq:pair_pot}) would seem a natural
choice for $v_{BB}(r)$. When one
considers the GCM to be a simple model for polymers in solution, then the
following empirical rules for the pair potential parameters apply between unlike
species: \cite{Dautenhahn,paper1,Archer1,Archer5} $R_{i \neq
j}^2=(R_{ii}^2+R_{jj}^2)/2$ and $\epsilon_{i \neq j} < \epsilon_{ii} \simeq
\epsilon_{jj}$. Therefore, the choices $R_{BB}/R_{11}=7$ and
$\epsilon_{BB}=2k_BT$ would be consistent with the parameters we have used for
the big-small pair potentials. \cite{Archer5} If we employ a bare Gaussian
potential with these parameters the big-big repulsion
is negligible when compared to the attractive $W_{BB}(r)$,
particularly when there are thick adsorbed wetting films present around the big
particles. Thus, the resulting effective pair potential, $v_{BB}^{eff}(r)$,
given by Eq.\ (\ref{eq:v_eff}) can be very strongly attractive.

If one were seeking to investigate experimentally the effects
of thick adsorbed films and
bridging between colloidal particles, one approach is to perform light
scattering experiments in order to measure the second virial coefficient, $B_2$.
\cite{deHek:VrijJColIntSci1982} As pointed out in Ref.~\onlinecite{Archer5}, $B_2$,
which measures the integral of $-r^2( \exp[-\beta v_{BB}^{eff}(r)]-1)$ should be
very large and negative when adsorbed films are present. A rapid change to large
negative values of $B_2$ upon changing the solvent state point should indicate
the development of thick adsorbed films around the colloids, thereby influencing
the SM potential. \cite{Archer5} Whether $B_2$ does show a rapid variation with
composition in the neighbourhood of the thin-thick transition lines remains to
be ascertained.

Finally we note that, just as a pair of big particles with $h=0$ exhibits a
thin-thick adsorbed film transition line, at total
densities higher than the single
particle thin-thick adsorbed transition line (see Fig.\ \ref{fig:phase_diag}),
there should also be a thin-thick adsorbed film
transition line at even higher densities for three big particles whose
centres coincide. This will have implications
for the three-body interactions between the big particles. Furthermore, there
may be other transition lines corresponding to four, five or more big particles
completely overlapping.

\section*{Acknowledgements}
We thank H. Stark for illuminating discussions and for providing a preprint of
Ref.~\onlinecite{Stark}. We also benefited from conversations with D. Andrienko and
O.I. Vinogradova.
AJA is grateful for the support of EPSRC under grant number GR/S28631/01. MO
thanks the Alexander von Humboldt Foundation  for making his stay in Bristol possible.

\end{document}